\documentclass[12pt,aps,print,showpacs]{revtex4-1}
\usepackage{amsfonts}
\usepackage{amsmath}
\usepackage{amssymb}
\usepackage{graphicx}
\usepackage{epstopdf}
\usepackage{subfigure}
\usepackage{hyperref}
\usepackage{array}
\usepackage{rotating}

\newcommand{\ket}[1]{|#1\rangle}

\begin{document}

\title{Fault-Tolerant Quantum Walks}
\author{S. D. Freedman}
\author{Y. Tong}
\author{J. B. Wang}
\email{jingbo.wang@uwa.edu.au}
\affiliation{School of Physics, The University of Western Australia, Perth, Australia}

\begin{abstract}
Quantum walks are expected to serve important modelling and algorithmic applications in many areas of science and mathematics. Although quantum walks have been successfully implemented physically in recent times, no major efforts have been made to combat the error associated with these physical implementations in a fault-tolerant manner.  In this paper, we propose a systematic method to implement fault-tolerant quantum walks in discrete time on arbitrarily complex graphs, using quantum states encoded with the Steane code and a set of universal fault tolerant matrix operations.
\end{abstract}

\maketitle


\section{Introduction}

Classical random walks have been used in the process of modelling natural systems in fields such as biology \cite{sessions97} and chemistry \cite{trautt06}. They have also been used in applied mathematics, in areas including optimization \cite{berg93} and solving differential equations \cite{hoshino71}. Quantum walks, the quantum analogue to classical random walks, will likely provide a more efficient way to carry out these processes, due to their non-intuitive and markedly different properties, including faster mixing and hitting times and resonance tunnelling, as well as the additional resource of interaction and entanglement between multiple quantum walkers \cite{kempe03, childs09, berry11, li13}. For these reasons, it is also expected that quantum walks will be employed in applications where it would be impractical or virtually impossible to employ classical random walks due to their relative inefficiency. Indeed, quantum walks have already given rise to efficient algorithms that have applications in graph theory, quantum search, network analysis, and other areas \cite{kempe03, berry11, Shenvi, douglas08, childs09, Reitzner, berry10, SmithMosca2012}.

This growing interest in the theoretical applications of quantum walks has inspired many experimentalists to implement quantum walks in various physical systems with state-of-the-art techniques, including nuclear magnetic resonance \cite{du03}, quantum optics \cite{zhang07, zhang10}, ion traps \cite{Schmitz2009, zahringer10}, neutral atom traps \cite{karski09}, linear optics \cite{Broome2010, pandey11, Schreiber2012}, and integrated photonics \cite{Peruzzo2010, Sansoni2012}.  Manouchehri and Wang \cite{manouchehripress} provided 
a comprehensive review on the numerous proposals for a physical realisation of quantum walks, as well as actual experiments, underpinned by a wide range of quantum, classical and hybrid technologies. 

In any experimental implementation of a quantum walk, error may be introduced due to the failure of any of the experiment's components, and this error can propagate throughout the experimental system. Therefore, it is vitally important for quantum walks to be implemented fault-tolerantly; that is, in a way that ensures that error due to the failure of a single component can be corrected and will not significantly affect the result of the experiment. However, there has so far been no major attempt to achieve this fault-tolerant implementation of quantum walks.  Additionally, most experimental implementations of quantum walks have so far been concerned with quantum walks on a 1D line or on simple symmetric graphs. This paper presents a method for fault-tolerantly implementing quantum walks in discrete time on any arbitrarily complex graphs.   

The paper is structured as follows. A very brief introduction to discrete-time quantum walks and the quantum walk operator $U$ is provided in Section~\ref{subsec:qw}. Section~\ref{subsec:csd} gives a description of how the quantum walk operator $U$ can be decomposed using the technique of cosine sine decomposition. An overview of the Steane code and a related universal set of fault-tolerant matrix operations, which are essential to the fault-tolerant implementation of quantum walks, is given in Section~\ref{subsec:steane}. Section~\ref{subsec:ftqw} describes how to implement a fault-tolerant quantum walk corresponding to an arbitrary $U$. In Section~\ref{sec:results}, we present our results, including an example of a quantum walk on the 8-star graph. Finally, Section~\ref{sec:conclusion} concludes, and provides suggestions for further research in this area.

\section{Theory}

\subsection{Discrete-time quantum walks}
\label{subsec:qw}

Consider an undirected graph $G(V,E)$, characterized by a set of vertices $V=\left\{v_1,v_2,v_3,\ldots\right\}$ and edges $E=\left\{(v_i,v_j),(v_k,v_l),\ldots\right\}$ consisting of connected vertex pairs. The state of a discrete-time quantum walker on such a graph is given by
\begin{equation}\label{eq:walkerstate}
\ket{\psi} = \sum_{(j,k) \in E}  \mathcal{A}_{j,k} \ket{j,k},
\end{equation}
where $\ket{j,k}$ and $\ket{k,j}$ are the two states corresponding to the vertex pair $(j,k)$, and $\mathcal{A}_{j,k}$ are complex amplitudes. If two particular nodes $j$ and $k$ are not connected by an edge, then $\mathcal{A}_{j,k} = 0$ and $\mathcal{A}_{k,j} = 0$.  The total number of states $N$ required in Eq.~\ref{eq:walkerstate} equals twice the number of edges between distinct vertices, plus the number of self loops, in the underlying graph.  At each step of the quantum walk, the walker's state undergoes the transformation
\begin{equation}\label{eq:walkerstatetransform}
\ket{\psi} \mapsto T\mathcal{C}\ket{\psi} \equiv U\ket{\psi},
\end{equation}
where $T$ is the conditional translation operator such that $T \ket{j,k} = \ket{k,j}$, $\mathcal{C}$ is the coin operator which mixes the complex amplitudes of groups of states associated with each vertex, and $U=T \mathcal{C}$ is the overall quantum walk operator.  Note that $T$, $\mathcal{C}$ and $U$ are all unitary operators.

\subsection{CS decomposition}
\label{subsec:csd}

Instead of implementing the $N\times N$ quantum walk operator $U$ directly, which would be difficult using any physical system, it can be replaced by a series of pairwise operators using the technique of cosine sine (CS) decomposition \cite{sutton09, manouchehri09, chen13}, for example. For this scheme to work, it is required that $U$ is an $\mathcal{N}\times\mathcal{N}$ matrix, where $\mathcal{N} = 2^M$, for some integer $M$. If $U$ is not already in this form, then $U$ undergoes the transformation
\begin{equation}\label{eq:UtransformCS}
U \mapsto \left(
\begin{array}{cc}
 U &  \\
 & I \\
\end{array}
\right).
\end{equation}
Here, $I$ is the $(\mathcal{N}-N)\times(\mathcal{N}-N)$ identity matrix, where $\mathcal{N}$ is the smallest power of $2$ larger than $N$. Correspondingly, the walker's state is transformed into a state containing $\mathcal{N}$ elements, the last $(\mathcal{N}-N)$ of which are initially $0$.

After recursively applying CS decomposition $M-1$ times to $U$, we find that
\begin{equation}\label{eq:csd}
U = \prod_{i=1}^{\mathcal{N}-1} \mathcal{U}_{i}(d_{i}), \text{where } \mathcal{U}_{i}(d_{i}) = \left(
\begin{array}{ccc}
 u_{i,1} &  & \\
  &  u_{i,2} &  \\
  &  & \ddots \\
\end{array}
\right).
\end{equation}
Here, each $u_{i,k}$ is a $d_{i} \times d_{i}$ block along the diagonal of $\mathcal{U}_{i}$, with $k = 1, 2, \ldots, \mathcal{N}/d_{i}$. Each $d_{i}$ can take a value in $\{ 2, 4, 8, \ldots, \mathcal{N}/2 \}$. When $d_{i} = 2$, $u_{i,k}$ is a $2 \times 2$ unitary matrix, and when $d_{i} > 2$,
\begin{equation}\label{eq:csblock}
u_{i,k} = \left(
\begin{array}{ccc|ccc}
 \ddots & & & \ddots & & \\
 & c_{r} & & &\hspace{8 pt} s_{r} & \\
 & & \ddots & & & \ddots \\
\hline
 \ddots & & & \ddots & & \\
 & -s_{r} & & & \hspace{8 pt}c_{r} & \\
 & & \ddots & & & \ddots \\
\end{array}
\right).
\end{equation}
Each quadrant of $u_{i,k}$ is diagonal, with entries $c_{r}$ and $s_{r}$ corresponding to cos$(\phi_{r})$ and sin$(\phi_{r})$, where $\phi_{r}$ is some angle, and $r = 1, 2, \ldots, d/2$.

If the walker's state in a quantum walk on a particular graph is given by
\begin{equation}\label{eq:statenum}
\ket{\psi} = \left(
\begin{array}{c}
a_{1} \\
a_{2} \\
\vdots \\
a_{\mathcal{N}} \\
\end{array}
\right),
\end{equation}
then $U \ket{\psi}$ can be further decomposed into a series of operations, each transforming two elements of $\ket{\psi}$, with the $m$th operation given by
\begin{equation}\label{eq:statedec}
\ket{\alpha_m} \equiv \left(
\begin{array}{c}
a_{p_{m}} \\
a_{q_{m}} \\
\end{array}
\right) \mapsto \mathcal{V}_{m}(\phi_{m})  \left(
\begin{array}{c}
a_{p_{m}} \\
a_{q_{m}} \\
\end{array}
\right).
\end{equation}
Here, $\ket{\alpha_m}$ consists of the two elements of $\ket{\psi}$ transformed by the $m$th operation, and each $\mathcal{V}_{m}(\phi_{m})$ is either a rotation matrix about the $y$ axis, 
\begin{equation}
R_y(\phi_{m}) = \left(
\begin{array}{cc}
\text{cos}(\phi_{m}) & \text{sin}(\phi_{m})   \\
-\text{sin}(\phi_{m})  &  \text{cos}(\phi_{m})   \\
\end{array}
\right),
\end{equation}
a rotation matrix about the $z$ axis, 
\begin{equation}
R_z(\phi_{m} )=\left(
\begin{array}{cc}
\text{exp}(i\phi_{m}) & 0   \\
0  &  \text{exp}(-i\phi_{m})   \\
\end{array}
\right),
\end{equation}
or a phase matrix, 
\begin{equation}
\Phi(\phi_{m}) = \left(
\begin{array}{cc}
1 & 0   \\
0  &  \text{exp}(i\phi_{m})   \\
\end{array}
\right).
\end{equation}
In the case of a real quantum walk operator $U$, each matrix $\mathcal{V}_{m}(\phi_{m})$ is either $R_y(\phi_{m})$ or $\Phi(\pi) \equiv Z$. 

\subsection{Fault-tolerant Steane code gates}
\label{subsec:steane}

Fault-tolerant gate operations can be implemented on qubits encoded using, for example, the 7-qubit error-correcting code developed by Steane \cite{steane96, nielsen10}, in a way that will allow us to implement fault-tolerant quantum walks as discussed in Section~\ref{subsec:ftqw}. Using this code, the qubit states $\ket{0}$ and $\ket{1}$ are respectively encoded as
\begin{equation}\label{eq:steane0}
\begin{split}
\ket{0_{L}} = \frac{1}{\sqrt{8}} (&\ket{0000000}+\ket{1010101}+\ket{0110011}+\ket{1100110}+ \\
&\ket{0001111}+\ket{1011010}+\ket{0111100}+\ket{1101001})
\end{split}
\end{equation}
and
\begin{equation}\label{eq:steane1}
\begin{split}
\ket{1_{L}} = \frac{1}{\sqrt{8}} (&\ket{1111111}+\ket{0101010}+\ket{1001100}+\ket{0011001}+ \\
&\ket{1110000}+\ket{0100101}+\ket{1000011}+\ket{0010110}).
\end{split}
\end{equation}
One universal set of single-qubit gates (denoted here as $\mathcal{G}$), which can be implemented fault-tolerantly when applied to qubits encoded using the Steane code, consists of the following elements \cite{nielsen10, fowler11}:
\begin{subequations}\label{eq:gates}
\begin{align}
&H = \left(
\begin{array}{cc}
 \frac{1}{\sqrt{2}} & \frac{1}{\sqrt{2}} \\
 \frac{1}{\sqrt{2}} & -\frac{1}{\sqrt{2}} \\
\end{array}
\right),\\
&X = \left(
\begin{array}{cc}
 0 & 1 \\
1 & 0 \\
\end{array}
\right),\\
&Z = \left(
\begin{array}{cc}
 1 & 0 \\
0 & -1 \\
\end{array}
\right),\\
&T = \left(
\begin{array}{cc}
 1 & 0 \\
 0 & \text{exp}(\frac{i \pi }{4}) \\
\end{array}
\right),\\
& S = T^2 = \left(
\begin{array}{cc}
 1 & 0 \\
 0 & i \\
\end{array}
\right),\\
& s = S^{\dagger} = \left(
\begin{array}{cc}
 1 & 0 \\
 0 & -i \\
\end{array}
\right).
\end{align}
\end{subequations}
Although the subset $\{H,T\}$ of $\mathcal{G}$ is also a universal set \cite{nielsen10}, using this subset is inefficient compared to the full set, as implementing the $T$ gate is much more complicated than implementing the other gates in $\mathcal{G}$ \cite{aliferis06}. Note that the gate $T^{\dagger}$ is not included in $\mathcal{G}$, as using this gate in gate sequences containing $2$ or more gates is equally as efficient as using the gate sequences $sT$ or $Ts$, which are equivalent to $T^{\dagger}$.

The fault-tolerant $H$, $X$ and $Z$ gates are implemented by applying regular $H$, $X$ or $Z$ gates respectively to each of the seven qubits corresponding to a single encoded qubit. The implementations of the fault-tolerant $S$ gate is similar, but the regular gate applied to each qubit in this case is the conjugate transposes of the fault-tolerant gate, $s$. It follows that the fault-tolerant $s$ gate can be implemented in the same way, applying the regular $S$ gate to each qubit.

The fault-tolerant T gate requires an input of two encoded qubits, i.e. two blocks of seven qubits \cite{nielsen10}. The first is prepared in the ancilla state
\begin{equation}\label{eq:ancilla}
\ket{\Theta} = \frac{\ket{0}+\text{exp}(\frac{i\pi}{4})\ket{1}}{\sqrt{2}},
\end{equation}
and the second is $\ket{\alpha}$, the qubit for which we desire the output $T\ket{\alpha}$. The ancilla state is prepared by first preparing the encoded $\ket{0_{L}}$ state, and then fault-tolerantly measuring exp$(-\frac{i\pi}{4})SX$, using the procedure described by Nielsen and Chuang \cite{nielsen10}. If the measurement gives a result of $+1$, then the ancilla state has been prepared. If the measurement result is $-1$, then the ancilla state preparation is completed by applying a fault-tolerant Z operation to the first qubit. Once the ancilla preparation is complete, a fault-tolerant CNOT gate is applied, with the first and second qubits as the control and target qubits respectively. This fault tolerant gate is implemented by applying a regular CNOT gate between each corresponding pair of single qubits in the two seven-qubit blocks. Next, the second qubit is measured fault-tolerantly, again using the method described by Nielsen and Chuang. If the result of the measurement is 0, then the first qubit is in the state $T\ket{\alpha}$, as required. Otherwise, applying the fault-tolerant $S$ and $X$ gates to the first qubit will transform the first qubit into the desired $T\ket{\alpha}$ state.

\subsection{Fault-tolerant quantum walks}
\label{subsec:ftqw}

For certain highly symmetric graphs, exponentially efficient quantum circuit implementation of the associated quantum walks are attainable \cite{brendan09, loke11, loke12}.  In these cases, the well-established fault-tolerant quantum computation protocols as discussed in \cite{Shor96, Steane97, PRESKILL98, nielsen10} can be readily applied.  
In what follows, however, we present a systematic method for a fault-tolerant implementation of discrete time quantum walks on any arbitrarily complex graph which, in general, cannot be implemented efficiently using a quantum circuit.  

By the Solovay-Kitaev theorem \cite{dawson06, nielsen10}, finite product sequences of matrices in the universal set $\mathcal{G}$ given in Eq.~\ref{eq:gates} can be used to approximate the matrices appearing in the decomposition of an arbitrary quantum walk operator $U$ acting on the state of a quantum walker $\ket{\psi}$.  The sequence length is in the order of $O(log^c(1/\epsilon))$, where $c$ is a small constant and $\epsilon$ is the desired precision. Since all matrices in the set $\mathcal{G}$ can be implemented fault-tolerantly as discussed in the previous section, so can the decomposition of $U\ket{\psi}$ that involves these matrices. In order to determine the fault-tolerant approximation of an operation in the decomposition of $U\ket{\psi}$, one needs to find a product sequence of matrices in $\mathcal{G}$ that approximates the matrix $\mathcal{V}_{m}(\phi_{m})$ corresponding to that operation. Then, the single operation is replaced by a series of operations, corresponding to the product sequence, each of which acts on the same state vector $\ket{\alpha_{m}}$. Note that in the case of a real quantum walk operator, only approximations of rotation matrices about the $y$ axis, i.e. real rotation matrices, need to be found. 

In order to find approximations of unitary matrices using product sequences of matrices in $\mathcal{G}$, Fowler \cite{fowler11} suggested finding a sequence, out of all sequences up to a certain length, that best approximates an individual matrix. With this approach, the matrix corresponding to a particular sequence is not checked against the desired matrix if the sequence is equivalent to a shorter sequence that has already been checked, and so this method is more efficient than checking all sequences exhaustively. However, since this method involves finding the approximation of a particular matrix, the procedure needs to be repeated every time the approximation of a new matrix is required. Therefore, 
a more efficient method, in terms of the required computational resources, is to search all possible sequences up to a certain length for those that sufficiently approximate any of the matrices $\mathcal{V}_{m}(\phi_{m})$. In order to further maximise efficiency, all sequences and corresponding matrices are stored so that all operations on sequences and matrices during the search are based on stored sequences of shorter length, and their corresponding matrices. Rules are also implemented to ensure that sequences are ignored if they are equivalent to sequences of the same or shorter length that have already been checked. Once all sequences up to a certain length have been checked, each matrix $\mathcal{V}_{m}(\phi_{m})$ can be associated with either a sequence that best approximates $\mathcal{V}_{m}(\phi_{m})$ out of all checked sequences, or a sequence of the shortest possible length that sufficiently approximates $\mathcal{V}_{m}(\phi_{m})$.


Since each two-level quantum state vector needs to be encoded as 7 qubits using the Steane code, and since each fault-tolerant $T$ gate requires an input of two 7-qubit blocks, each operation of an element in $\mathcal{G}$ on $\ket{\alpha_{m}}$ in the final fault-tolerant approximation of the decomposition of $U\ket{\psi}$ involves a $2^{14}\times2^{14}$ matrix, acting on $2^{14}$ elements of the walker's encoded state vector. If the quantum walk involving these fault-tolerant operations is implemented physically, then given that the probability of error arising due to any single component of the implementation is sufficiently small, any error introduced can be effectively corrected \cite{nielsen10}. In this way, the results of this fault-tolerant quantum walk can be made arbitrarily close to those obtained by implementing the original quantum walk with no error involved.

\section{Results}
\label{sec:results}

Product sequences of matrices in $\mathcal{G}$ of length up to $37$ were checked, using the more efficient method described in Section~\ref{subsec:ftqw}, to find sequences that approximate real rotation matrices. A matrix corresponding to a sequence was accepted as an approximate real rotation matrix of angle $\phi$ if it was in the form of
\begin{equation}\label{eq:approxdef}
\left(
\begin{array}{cc}
\text{cos}(\phi) + ai & \text{sin}(\phi\pm\varepsilon) + bi \\
-\text{sin}(\phi\pm\varepsilon) + ci & \text{cos}(\phi) + di \\
\end{array}
\right),
\end{equation}
with the condition $|a|,|b|,|c|,|d| < 0.1$. Here, $\varepsilon$ is an error that was accepted as small enough if the aforementioned condition was met. Sequences corresponding to $1213$ separate approximate angles between $0^{\circ}$ and $180^{\circ}$ inclusive were found, with an average difference of $0.149^{\circ}$ between successive angles. The list of angles found included only $9$ successive angle differences greater than $1^{\circ}$, the largest being a difference of $2.459^{\circ}$, and the smallest difference between successive angles on the list was $0.052^{\circ}$. Additionally, if a sequence of a certain length was found to approximate a rotation matrix of angle $\phi$, then another sequence of the same length was found to approximate a rotation matrix of angle $-\phi$, to the same accuracy. 

\begin{sidewaysfigure}[h]
	\centering
	\vspace{0cm} \hspace{0cm} \scalebox{0.96}{\includegraphics{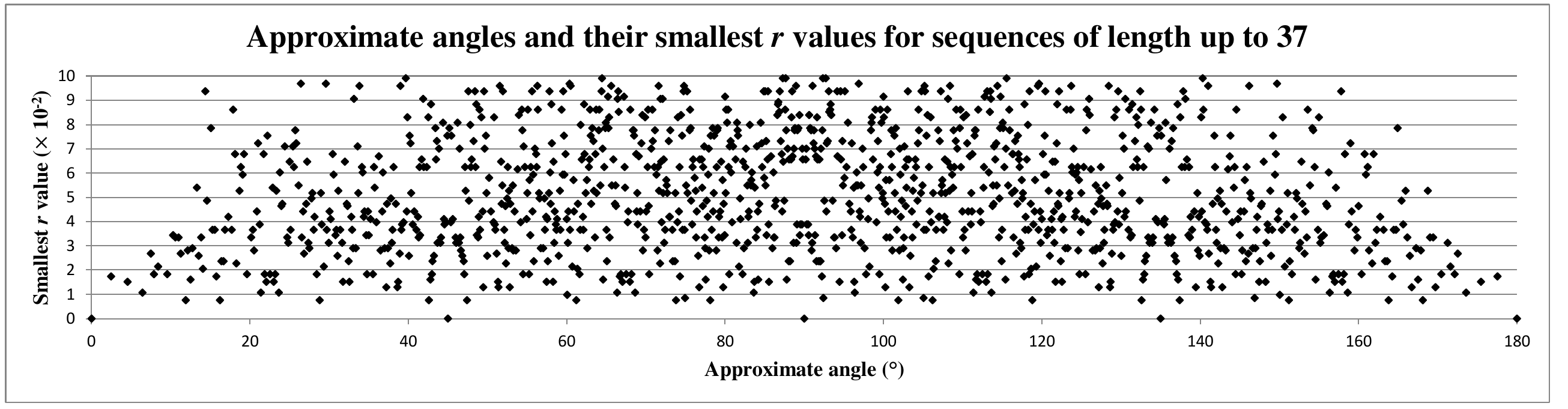}}
	\caption{All approximate angles found, and their corresponding smallest $r$ values, for sequences of length up to 37.}\label{fig:angles}
\end{sidewaysfigure}

\autoref{fig:angles} is a plot of all approximate angles found, together with their smallest respective $r$ values, where $r = \text{max}\{|a|,|b|,|c|,|d|\}$. A file ($AngleSequenceTables.xlsx$) containing two pairs of tables of all angles found, 
together with a sequence of length up to 37 that approximates the rotation matrix corresponding to each angle, is located at ftp://ftp.physics.uwa.edu.au/pub/QuantumComputation. In one table of each pair, the sequence for each angle is chosen first by smallest $r$ value and then by shortest length, and in the other, the sequence is chosen first by shortest length and then by smallest $r$ value. One pair of tables corresponds to positive angles, and the other to negative angles. Negative angles are /d as the corresponding sequences cannot be obtained by simply inverting the sequences corresponding to positive angles, since the $T^{\dagger}$ gate is not included in $\mathcal{G}$, as discussed in Section~\ref{subsec:steane}. Although each table also includes an angle of $0^{\circ}$, the corresponding sequence is not included in any decomposition of $U\ket{\psi}$, as the rotation matrix of angle $0^{\circ}$ is the identity matrix.

\begin{figure}[h]
	\centering
	\vspace{0cm} \hspace{0cm} \scalebox{1.0}{\includegraphics{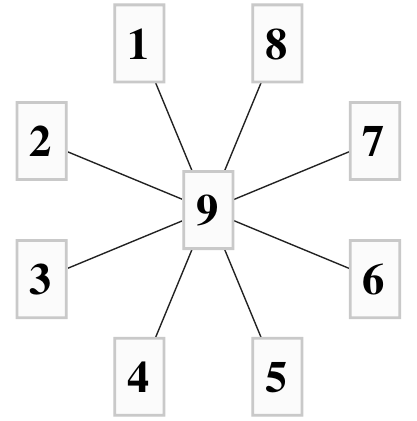}}
	\caption{The 8-star graph.}\label{fig:8star}
\end{figure}

We now consider, as an example, a quantum walk on the 8-star graph shown in \autoref{fig:8star}.
The quantum walk operator $U$ corresponding to this graph, using the Grover Coin, is the real $16\times16$ matrix \cite{chen13},
\begin{equation}\label{eq:inputmat}
U=\left(
\begin{array}{cccccccccccccccc}
 0 & 0 & 0 & 0 & 0 & 0 & 0 & 0 & -0.75 & 0.25 & 0.25 & 0.25 & 0.25 & 0.25 & 0.25 & 0.25 \\ [-1.5ex]
 0 & 0 & 0 & 0 & 0 & 0 & 0 & 0 & 0.25 & -0.75 & 0.25 & 0.25 & 0.25 & 0.25 & 0.25 & 0.25 \\ [-1.5ex]
 0 & 0 & 0 & 0 & 0 & 0 & 0 & 0 & 0.25 & 0.25 & -0.75 & 0.25 & 0.25 & 0.25 & 0.25 & 0.25 \\ [-1.5ex]
 0 & 0 & 0 & 0 & 0 & 0 & 0 & 0 & 0.25 & 0.25 & 0.25 & -0.75 & 0.25 & 0.25 & 0.25 & 0.25 \\ [-1.5ex]
 0 & 0 & 0 & 0 & 0 & 0 & 0 & 0 & 0.25 & 0.25 & 0.25 & 0.25 & -0.75 & 0.25 & 0.25 & 0.25 \\ [-1.5ex]
 0 & 0 & 0 & 0 & 0 & 0 & 0 & 0 & 0.25 & 0.25 & 0.25 & 0.25 & 0.25 & -0.75 & 0.25 & 0.25 \\ [-1.5ex]
 0 & 0 & 0 & 0 & 0 & 0 & 0 & 0 & 0.25 & 0.25 & 0.25 & 0.25 & 0.25 & 0.25 & -0.75 & 0.25 \\ [-1.5ex]
 0 & 0 & 0 & 0 & 0 & 0 & 0 & 0 & 0.25 & 0.25 & 0.25 & 0.25 & 0.25 & 0.25 & 0.25 & -0.75 \\ [-1.5ex]
 1 & 0 & 0 & 0 & 0 & 0 & 0 & 0 & 0 & 0 & 0 & 0 & 0 & 0 & 0 & 0 \\ [-1.5ex]
 0 & 1 & 0 & 0 & 0 & 0 & 0 & 0 & 0 & 0 & 0 & 0 & 0 & 0 & 0 & 0 \\ [-1.5ex]
 0 & 0 & 1 & 0 & 0 & 0 & 0 & 0 & 0 & 0 & 0 & 0 & 0 & 0 & 0 & 0 \\ [-1.5ex]
 0 & 0 & 0 & 1 & 0 & 0 & 0 & 0 & 0 & 0 & 0 & 0 & 0 & 0 & 0 & 0 \\ [-1.5ex]
 0 & 0 & 0 & 0 & 1 & 0 & 0 & 0 & 0 & 0 & 0 & 0 & 0 & 0 & 0 & 0 \\ [-1.5ex]
 0 & 0 & 0 & 0 & 0 & 1 & 0 & 0 & 0 & 0 & 0 & 0 & 0 & 0 & 0 & 0 \\ [-1.5ex]
 0 & 0 & 0 & 0 & 0 & 0 & 1 & 0 & 0 & 0 & 0 & 0 & 0 & 0 & 0 & 0 \\ [-1.5ex]
 0 & 0 & 0 & 0 & 0 & 0 & 0 & 1 & 0 & 0 & 0 & 0 & 0 & 0 & 0 & 0 \\ [-1.5ex]
\end{array}
\right).
\end{equation}
This operator, acting on the state of a quantum walker, can be decomposed as described in Eq.~\eqref{eq:statedec}, and then each matrix in the decomposition can be replaced by product sequences of matrices in $\mathcal{G}$. Table~\ref{table:angtable} lists the matrices $\mathcal{V}_{m}(\phi_{m})$ in the decomposition of $U\ket{\psi}$, and the elements of $\ket{\psi}$ that they apply to. Table~\ref{table:seqtable} describes the operations in the fault-tolerant decomposition of $U\ket{\psi}$. Specifically, each rotation matrix of angle $\phi$ in Table~\ref{table:angtable} is replaced with the product sequence of matrices in $\mathcal{G}$, out of all such sequences of length up to $37$, that most closely approximates a rotation matrix whose angle is as close as possible to $\phi$. 
In each of these tables, the operations corresponding to the first two columns appear in the overall sequence of operations before those corresponding to the last two columns, with operations later in the overall sequence being applied first, as in the case of a sequence of matrices acting on a column vector. A total of $763$ matrices in $\mathcal{G}$ appear in this fault-tolerant decomposition of $U\ket{\psi}$.

Applying the operations described in Table~\ref{table:seqtable} to $\ket{\psi}$, directly using the matrices in $\mathcal{G}$ instead of the corresponding fault-tolerant $2^{14}\times2^{14}$ matrices, is equivalent to applying a complex unitary matrix to $\ket{\psi}$ whose real part is
\begin{equation}\label{eq:apmat}
\left(
\begin{array}{cccccccccccccccc}
 0 & 0 & 0 & 0 & 0 & 0 & 0 & 0 & -0.7529 & 0.2532 & 0.2436 & 0.2426 & 0.2461 & 0.2466 & 0.2470 & 0.2468 \\ [-1.5ex]
 0 & 0 & 0 & 0 & 0 & 0 & 0 & 0 & 0.2533 & -0.7374 & 0.2515 & 0.2411 & 0.2516 & 0.2520 & 0.2524 & 0.2522 \\ [-1.5ex]
 0 & 0 & 0 & 0 & 0 & 0 & 0 & 0 & 0.2433 & 0.2516 & -0.7243 & 0.2351 & 0.2493 & 0.2499 & 0.2503 & 0.2501 \\ [-1.5ex]
 0 & 0 & 0 & 0 & 0 & 0 & 0 & 0 & 0.2428 & 0.2411 & 0.2350 & -0.7195 & 0.2491 & 0.2501 & 0.2505 & 0.2503 \\ [-1.5ex]
 0 & 0 & 0 & 0 & 0 & 0 & 0 & 0 & 0.2467 & 0.2521 & 0.2499 & 0.2499 & -0.7291 & 0.2471 & 0.2466 & 0.2464 \\ [-1.5ex]
 0 & 0 & 0 & 0 & 0 & 0 & 0 & 0 & 0.2466 & 0.2520 & 0.2499 & 0.2501 & 0.2458 & -0.7417 & 0.2543 & 0.2361 \\ [-1.5ex]
 0 & 0 & 0 & 0 & 0 & 0 & 0 & 0 & 0.2463 & 0.2517 & 0.2495 & 0.2495 & 0.2438 & 0.2585 & -0.7442 & 0.2365 \\ [-1.5ex]
 0 & 0 & 0 & 0 & 0 & 0 & 0 & 0 & 0.2464 & 0.2518 & 0.2496 & 0.2495 & 0.2420 & 0.2374 & 0.2449 & -0.7247 \\ [-1.5ex]
 1 & 0 & 0 & 0 & 0 & 0 & 0 & 0 & 0 & 0 & 0 & 0 & 0 & 0 & 0 & 0 \\ [-1.5ex]
 0 & 1 & 0 & 0 & 0 & 0 & 0 & 0 & 0 & 0 & 0 & 0 & 0 & 0 & 0 & 0 \\ [-1.5ex]
 0 & 0 & 1 & 0 & 0 & 0 & 0 & 0 & 0 & 0 & 0 & 0 & 0 & 0 & 0 & 0 \\ [-1.5ex]
 0 & 0 & 0 & 1 & 0 & 0 & 0 & 0 & 0 & 0 & 0 & 0 & 0 & 0 & 0 & 0 \\ [-1.5ex]
 0 & 0 & 0 & 0 & 1 & 0 & 0 & 0 & 0 & 0 & 0 & 0 & 0 & 0 & 0 & 0 \\ [-1.5ex]
 0 & 0 & 0 & 0 & 0 & 1 & 0 & 0 & 0 & 0 & 0 & 0 & 0 & 0 & 0 & 0 \\ [-1.5ex]
 0 & 0 & 0 & 0 & 0 & 0 & 1 & 0 & 0 & 0 & 0 & 0 & 0 & 0 & 0 & 0 \\ [-1.5ex]
 0 & 0 & 0 & 0 & 0 & 0 & 0 & 1 & 0 & 0 & 0 & 0 & 0 & 0 & 0 & 0 \\ [-1.5ex]
\end{array}
\right). 
\end{equation}

Comparing this with $U$ in Eq.~\eqref{eq:inputmat}, it can be seen that the largest absolute error between corresponding elements is $0.0305$, and that the largest relative error between corresponding elements is $6.00\%$. Additionally, the largest magnitude of an element of the imaginary part of the approximation of $U$, which is only nonzero in its upper-right quadrant, is $0.180$.  The distance between a $w \times w$ unitary matrix $W$ and its approximation $W_{l}$ is given by the metric \cite{fowler11}
\begin{equation}\label{eq:metric}
d(W,W_{l}) = \sqrt{\frac{w-|\text{tr}(W^{\dagger}W_{l})|}{w}}.
\end{equation}
Using this metric, which can take values between $0$ and $1$, the distance between $U$ and its approximation is $0.0901$. This approximation is therefore very close to $U$, and so if a quantum walk is implemented using the operations described in
Table~\ref{table:seqtable}, the results of the walk will be approximately those obtained by implementing a quantum walk 
on the 8-star graph using $U$ with no error involved, given that the fault-tolerant fault-tolerant set $\mathcal{G}$ works effectively.

It is also possible, instead of choosing the sequence that best approximates a rotation matrix of a certain approximate angle, 
to choose the shortest sequence that approximates the rotation matrix, again from all sequences of length up to $37$. In this case, comparing the real part of the approximation of $U$ and $U$ itself, the largest absolute and relative errors between corresponding elements are $0.0671$ and $10.8\%$ respectively, and the largest magnitude of an element of the imaginary part of this approximation, which is again only nonzero in its upper-right quadrant, is $0.317$. Using the metric given in Eq.~\eqref{eq:metric}, the distance between $U$ and this approximation is $0.121$. This approximation is therefore less accurate than the previous one, and requires $735$ fault-tolerant matrices, which is over $96\%$ of the number of matrices required in the previous approximation. This indicates that longer sequences corresponding to better approximations should be used, at least in the case of this example.

\section{Conclusion}
\label{sec:conclusion}

Quantum walks will likely play an important role in modelling and algorithmic applications in science and mathematics, but their physical implementations will only be practical when the associated experimental error can be effectively corrected. In this paper, we have presented a general method of implementing a fault-tolerant quantum walk in discrete time, using quantum states encoded with the Steane code, that approximates a quantum walk on an arbitrarily complex graph. If such a quantum walk is physically implemented, then error introduced during the implementation can be corrected effectively, as long as the probability of error associated with each component of the implementation is sufficiently small. The provided example of a quantum walk on the 8-star graph indicates that the proposed scheme can be implemented successfully. In other words, the results obtained by implementing the fault-tolerant quantum walk can be made arbitrarily close, by choosing long enough sequences of matrices in $\mathcal{G}$, to those obtained in the ideal case of implementing the original quantum walk with zero experimental error.

Tables of the sequences of matrices in $\mathcal{G}$ that approximate real rotation matrices, essential to the fault-tolerant implementation of quantum walks involving both real and complex quantum walk operators, have been constructed for sequences of length up to 37. Further research in this area will focus on constructing similar tables of sequences that approximate rotation matrices $R_z(\phi_{m})$ and phase matrices $\Phi(\phi_{m})$, 
which will be required for the implementation of fault-tolerant quantum walks involving complex quantum walk operators.  Additionally, the developed tables can be extended to include sequences of longer lengths, in order to improve the accuracy to which arbitrary quantum walks can be fault-tolerantly approximated.

\newpage

\newpage

\clearpage
\begin{table}[h]
\caption{The matrices, and the corresponding elements in the quantum walker's state, in the decomposition of $U\ket{\psi}$. This table was generated using a modified version of the \emph{Qcompiler} package \cite{chen13}.}
\centering
\begin{tabular}{>{$}c<{$} |>{$}c<{$} |>{$}c<{$} |>{$}c<{$}}
\hline\hline
\text{Matrix} & p,q & \text{Matrix} & p,q \\ [0.5ex]
\hline
R_y(-90.000^{\circ}) & 1,9 & R_y(-50.057^{\circ}) & 11,12 \\
R_y(-90.000^{\circ}) & 2,10 & R_y(-64.471^{\circ}) & 13,14 \\
R_y(-90.000^{\circ}) & 3,11 & R_y(45.000^{\circ}) & 15,16 \\
R_y(-90.000^{\circ}) & 4,12 & R_y(-90.000^{\circ}) & 12,16 \\
R_y(-90.000^{\circ}) & 5,13 & R_y( 3.126^{\circ}) & 9,10 \\
R_y(-90.000^{\circ}) & 6,14 & R_y(50.057^{\circ}) & 11,12 \\
R_y(-90.000^{\circ}) & 7,15 & R_y(64.471^{\circ}) & 13,14 \\
R_y(-90.000^{\circ}) & 8,16 & R_y(45.000^{\circ}) & 15,16 \\
R_y(18.933^{\circ}) & 9,10 & R_y(-23.908^{\circ}) & 9,11 \\
R_y(-52.595^{\circ}) & 11,12 & R_y(81.630^{\circ}) & 10,12 \\
R_y( 22.500^{\circ}) & 13,14 & R_y(-22.500^{\circ}) & 13,15 \\
R_y(-22.500^{\circ}) & 15,16 & R_y(67.500^{\circ}) & 14,16 \\
R_y(23.908^{\circ}) & 9,11 & R_y(-18.933^{\circ}) & 9,10 \\
R_y(-81.630^{\circ}) & 10,12 & R_y(52.595^{\circ}) & 11,12 \\
R_y(22.500^{\circ}) & 13,15 & R_y(-22.500^{\circ}) & 13,14 \\
R_y(67.500^{\circ}) & 14,16 & R_y(-22.500^{\circ}) & 15,16 \\
R_y(-3.126^{\circ}) & 9,10 & Z & 15,16 \\ [1ex]
\hline
\hline
\end{tabular}
\label{table:angtable}
\end{table}

\begin{sidewaystable}
\caption{The matrix product sequences, and the elements in the quantum walker's state, in the approximation of the decomposition of $U\ket{\psi}$. This table was generated using a modified version of the \emph{Qcompiler} package \cite{chen13}.}
\centering
\begin{tabular}{>{$}c<{$} |>{$}c<{$} |>{$}c<{$} |>{$}c<{$}}
\hline\hline
\text{Sequence} & p,q & \text{Sequence}  &p,q \\ [0.5ex]
\hline
XZ & 1,9  & THTHTHTsHTHTHTsHTHTHTsHTHTHTHsXTHTHT & 11,12  \\ [-1ex]
XZ & 2,10  & sHTHTHTHTHSTHTHTsHTHTHTHTsHT & 13,14  \\ [-1ex]
XZ & 3,11  & ZH & 15,16  \\ [-1ex]
XZ & 4,12  & XZ & 12,16  \\ [-1ex]
XZ & 5,13  &  ZTHTHTHZTHTHTHTHTHTHTHTsHTHTHTHTHTHsX & 9,10  \\ [-1ex]
XZ & 6,14  &  THsXTHTHTHTHTHTsHTHTHTsHTHTHTsHTHTHT & 11,12  \\ [-1ex]
XZ & 7,15  &  THTsHTHTHTHTsHTHSTHTHTHTHTHs & 13,14  \\ [-1ex]
XZ & 8,16  &  ZH & 15,16  \\ [-1ex]
THSTHTHTHSTHTHTHTHTXZHTHTHSTHTHT & 9,10  & THTHTHTHTHZTHTHTHSTHTsHTHTHTHZTHTHTH & 9,11  \\ [-1ex]
sHZTHTHTHTHTsHTHTHSTHTHSTHTHT  & 11,12 & THTsHTHTHTHTHTHTsHTHTHTsHTHTHTHTsHTH  & 10,12 \\ [-1ex]
sHsXTHTHTHTsHTHTHSTHTHTHSTHTHTHTHT & 13,14  & sHTsHTHTHTsHTHTHSTHTHTHSTHTHTHTHTX & 13,15  \\ [-1ex]
sHTsHTHTHTsHTHTHSTHTHTHSTHTHTHTHTX & 15,16 & THTHTHZTHTHSTHTHTHTsHTsHTHSTHTHTHs & 14,16  \\ [-1ex]
HTHTHZTHTHTHTHTsHSTHTHTHZTHTHTHTHTHT & 9,11  & THTXZHTHTHSTHTHTHTHTHSTHTHTHSTHT & 9,10  \\ [-1ex]
HTHTsHTHTHTHTsHTHTHTsHTHTHTHTHTHTsHT & 10,12  & THTHSTHTHSTHTHTHTsHTHTHTHZTHs & 11,12  \\ [-1ex]
sHsXTHTHTHTsHTHTHSTHTHTHSTHTHTHTHT & 13,15  & sHTsHTHTHTsHTHTHSTHTHTHSTHTHTHTHTX & 13,14  \\ [-1ex]
THTHTHZTHTHSTHTHTHTsHTsHTHSTHTHTHs & 14,16   & sHTsHTHTHTsHTHTHSTHTHTHSTHTHTHTHTX & 15,16  \\ [-1ex]
THTHTHZTHTHTHTHTHTHTHTsHTHTHTHTHTHsXZ & 9,10  & Z & 15,16  \\ [1ex]
\hline
\hline
\end{tabular}
\label{table:seqtable}
\end{sidewaystable}

\bigskip

\clearpage

\bibliography{Draft}

\end{document}